\documentclass[12pt]{article}
\usepackage[latin1]{inputenc}
\usepackage{latexsym}
\usepackage{graphicx}
\usepackage{epsfig}

\def\.{\cdot}

\def\s{\sigma}

\def\o{\over}

\def\a{\alpha}

\def\d{\delta}

\def\p{\partial}

\def\na{\nabla}
\def\+{\bigoplus}

\def\({\left(}
\def\){\right)}
\def\[{\left[}
\def\]{\right]}
\def\l.{\left.}
\def\r.{\right.}

\def\be{\begin{equation}}
\def\ee{\end{equation}}
\def\bea{\begin{eqnarray}}
\def\eea{\end{eqnarray}}
\def\nn{\nonumber \\ &&}

\def\acca{\right.\nn\left.}
\def\ber{\begin{array}}
\def\eer{\end{array}}

\def\L{{\cal L}}

\begin{document}

\title{Slavnov-Taylor and Ward Identities in the Electroweak Theory\footnote{Paper written written  on the occasion of Andrey Slavnov's seventy fifth birthday}}

\author{C.  BECCHI\\
\\ Universit\`a di Genova, Dipartimento di Fisica\\and I.N.F.N. Sezione di Genova\\ via Dodecaneso 33, Genova\\I-16146, Italy
\\
E-mail:  becchi@ge.infn.it}
 \maketitle
 \thispagestyle{empty}
 
\begin{abstract}
In the framework of the Electroweak Theory we discuss a class of gauge fixing choices suitable for the calculation of electromagnetic processes. We show in particular that,  with our choices, beyond the basic Slavnov-Taylor identities guaranteeing the independence of the physical results of the particular  gauge fixing, one also has the standard QED Ward identities  which play a well known crucial role in the calculation of electromagnetic processes and, in particular, in the analysis of the electromagnetic radiative corrections.
\end{abstract}




\section{Introduction}

Undergraduate students in Elementary Particle Physics know that the main aim of the Electroweak Theory is the unitarization of the Fermi theory of weak interactions. This is based on  the introduction of intermediate vector bosons which must be charged, and hence coupled to the electromagnetic field, and massive, since the beta decay spectra
show that Fermi's interaction has short range. The Electroweak Theory is based upon this dynamical scheme realized in the framework of renormalizable interactions, thus guaranteing perturbative unitarity. The theory is based on  the Yang-Mills  model of non-Abelian gauge interactions whose perturbative consistency with massive vector bosons  has been proved in the period 1963-67 under the condition that a new neutral scalar particle, named the Higgs particle,  exists. The discovery of the neutral weak currents, followed by that of the intermediate massive vector bosons and by the very recent one of the Higgs particle, have put the Electroweak Theory on the most solid bases \cite{bri}.

The above paragraph is an extremely short description of the kernel of the Electroweak Theory, still there are basic technical aspects of the theory, which are less known to university students, but  are essential for its consistency. Infact, in a Lorentz invariant framework, the renormalizability of the theory  requires the introduction, together with new physical particles, of unphysical ones associated with fields, or, better, field components, which are required by symmetry conditions, or else, by the unitarity relations. Among these fields there are the scalar components of the vector bosons, required by explicit Lorentz invariance, three more components of the Higgs field, required by the weak isotopic invariance, and the, much more tricky, Faddeev-Popov ghosts, which must compensate the contributions of the other unphysical fields to the unitarity relations.

The technical problem associated with the unphysical content of the theory is due to the lack of uniqueness of the construction. Giving up  explicit Lorentz invariance, one could avoid introducing scalar components of the vector fields and hence also reduce the role of the other unphysical fields. However one has to  guarantee the Lorentz invariance of the physical results, which is by no means a trivial fact. A lack of uniqueness also exists  in Feynman's QED, however, at least in the commonly considered case of linear gauge fixing conditions,  electromagnetic current conservation, which is a physical property,   guarantees that the scalar component of the electromagnetic vector potential be a free field, and hence, one can disregard it, although it appears in the Lagrangian of the theory. At the quantum level this is made explicit by the well known QED Ward identities.
It is well known that Ward identities play a crucial  role in the calculations of  electromagnetic processes and of, real and virtual, radiative corrections, as for example, in the proof of their factorization \cite{yfs}.

However, since the weak vector boson are charged, the Electroweak Theory is a non-Abelian gauge theory. This implies that, in an explicit Lorentz invariant  treatment, the scalar components of the vector bosons are not anymore free fields and their couplings depend on the choice of the gauge fixing conditions. It was suggested by 't Hooft \cite{gth} and completely understood by Slavnov \cite{aas} and Taylor \cite{jct}, that the fulfillment of the unitarity conditions and the independence of the results of the gauge fixing prescriptions involve a precise relation between  the dynamics of the unphysical field components and that of the Faddeev-Popov ghosts. Slavnov-Taylor relations are a basic property of the non-abelian gauge theories and essentially solve  the problems of unitarity and gauge independence.

Still one should not forget that in the early sixties, when 50 GeV center of mass energies were beyond technical reach, the most urgent problem was  computing the electromagnetic radiative corrections to the weak processes. With this, partially out of fashion, aim, it would be important to have recourse to the QED Ward identities which, however, do not hold true if the theory is quantized with the most frequently used gauge fixing prescriptions. More recently the calculation of high energy electromagnetic processes, such as the Higgs boson into two gammas decay \cite{egn}, have shown the advantages of schemes satisfying the QED Ward identities \cite{svvz} \cite{bri}. The purpose of the present note is to present a family of gauge fixing choices which guarantee the fulfillment of both, Slavnov-Taylor, and QED Ward identities.

It is worth recalling here the alternative  possibility of  exploiting background gauge choices. In the presence of  background gauge fields it is possible to ask for a theory  being invariant under gauge transformations of the background fields. It is known that this guarantees that the transition amplitudes involving gauge fields satisfy a non-abelian generalization of the QED Ward identities. A clear example of the advantages of this procedure has been given by Shifman et al. in the one-loop calculation of the Higgs boson decay amplitude into two gammas, where photons only appear in the final state \cite{svvz}.
What we have in mind is different, since we want to profit of the QED Ward identities also in the case of virtual photons.  Virtual and real photon processes mix together in the calculation of radiative corrections to the cross sections.

Therefore our purpose is  the construction of a Lagrangian density  invariant under electromagnetic gauge transformations,   with the exception of  the Feynman gauge fixing term. In the standard approaches
the gauge fixing part in general breaks every gauge invariance, however it is possible to keep electromagnetic gauge invariance when fixing the unphysical components of the weak intermediate vector bosons, $W^\pm$ and $Z$, using  electromagnetic  covariant gauge fixing functions. This is a consequence of the fact that the BRST operator ${\cal S}$, which appears in the construction of the Faddeev-Popov term, transforms electromagnetic  covariant quantities into electromagnetic  covariant ones. The unphysical components of the electromagnetic vector potential remain to be fixed. This can be done having recourse to  Feynman's prescription, which decouples the scalar component of the vector potential, if the rest of the Lagrangian is electromagnetic  gauge invariant. 

It is clear that our work will be purely technical, indeed the field tensor associated with the massless component of the intermediate vector bosons is not a physically well defined quantity; its matrix elements depend on the gauge fixing choice. The physically well defined tensor field is that built from the Abelian component of the gauge fields, which is a mixture of the neutral components\footnote{Using the notations of next section the physical field tensor corresponds, at the classical level, to $F^{\mu\nu}-\tan\theta_W Z^{\mu\nu}$.}. However, considering the low energy electromagnetic effects, the role of the massless component becomes dominant and hence it is worth profiting of the  QED Ward identities concerning these components.

We shall see that the cost of the QED Ward identities is the non-linearity of the $W_\pm$ gauge fixing functions, which become composite operators due to the presence of the electromagnetic covariant derivatives. Therefore, contrary to the case of linear gauge fixing, the parameters are affected with perturbative corrections to be controlled by suitable renormalization prescriptions. These must concern the coefficients of all possible terms of the gauge fixing function. Now the question  is, how high is this price. The answer depends on the purpose of the calculations. If one is considering high order corrections in the weak parameters, which might be necessary in a precision calculation of a very high energy process, the gauge fixing renormalization effects might become prohibitive. However, if, at the same time, one must take into account the low energy electromagnetic radiative corrections, the choice of our scheme might become compulsory. If, instead, one is willing to compute a one-loop weak process, such as the Higgs into two gamma decay the advantages of our scheme are the same as those of the background method and consist in a sharp reduction of the contributing diagrams.

\section{The Lagrangian}

Even if the Lagrangian and the Feynman rules of the Electroweak Theory are well known, at least in the restricted 't Hooft gauge ($\xi=1$), we present here the whole Lagrangian in the general  't Hooft $\xi$-gauge, in order that the reader  be able to compare our proposal with the schemes he is familiar with. We begin  presenting  the gauge invariant part of the Lagrangian density of the
standard model with one Higgs doublet. We disregard spinor fields since, in renormalizable frameworks, their dynamical properties are independent of the gauge fixing. 

In view of the particular role we  assign to the electromagnetic interactions, we denote by $\na$ the electromagnetic covariant derivative, that is, we define, for every charged field $\Phi_\pm$:
\be \na^\mu\Phi_\pm\equiv\p^\mu\Phi_\pm\mp ie A^\mu\Phi_\pm\ .\ee

The rules of the game are well known, we list them for completeness. The parameters are $g$, $g'$ and $\theta_W=\arctan (g'/g)$, the elementary electromagnetic charge is $e=g\sin\theta_W$.

The electromagnetic field tensor is $F^{\mu\nu}$; we set $Z^{\mu\nu}=\p^\mu Z^\nu-\p^\nu Z^\mu\ .$

The Higgs field involves the  doublet scalar field with components:
$$
\phi=\(\matrix {G^+ \cr\frac{H+v+iG}{\sqrt{2}}\cr}\)\quad
\,{\rm and\, its\, Hermitian\, conjugate}\quad
\phi^\dag=
\(\matrix {G^- \cr  \frac{H+v-iG}{\sqrt{2}}\cr}\)\ .
$$

The Lagrangian of the theory is the sum of a gauge  invariant part and of a gauge fixing part.
The gauge invariant part is the sum of two terms
\be
\L_{\rm SM}=\L_{\rm V}+\L_{\rm Higgs}
\ee
where  $\L_{\rm V}$ is the invariant part of  the gauge vector boson Lagrangian:
\bea
\L_{\rm V}&=&-\na_\mu W_{+\nu}\na_\mu W_{-\nu}+\na_\mu W^\mu_{+ }\na_\nu W^\nu_{- } -{1\o4}\[F_{\mu\nu} F^{\mu\nu} +Z_{\mu\nu} Z^{\mu\nu} \]\nn+2i\[eF_{\mu\nu} +g\cos\theta_WZ_{\mu\nu}\]W_+^\mu W_-^\nu\nn
 +ig\cos\theta_WZ_\mu\[W_{-\nu}\na^\mu W_+^\nu-W_{+\nu}\na^\mu W_-^\nu +W^\mu_-\na\cdot W_+ -W_+^\mu\na\cdot  W_-\]\nn
+{g^2\o2}\[  W_+^2W_-^2-(W_+\cdot W_-)^2\acca-\cos^2\theta_W\(Z^2(W_+\cdot W_-)-(Z\cdot W_+)(Z\cdot W_-)\)\]\ .\label{iv}
\eea
 and $\L_{\rm Higgs}$ is the invariant part of the Higgs Lagrangian. Setting $m_W=gv/2, \ m_Z=gv/(2\cos\theta_W)$ one has:
\bea
\L_{\rm Higgs}&=&{1\o2}\[(\p H)^2 +(\p G)^2\] +\na G_+\cdot\na G_-+
-{m_H^2\o2}\[(H+{H^2\o2v})^2\acca+{G_+G_-\o v}(H+{H^2\o2v}+{G^2+G_+G_-\o v})+{G^2\o v} (H+{H^2\o2v} +{G^2\o4v^2})\]
\nn
\left(m_W^2 {W^\mu}_+W_{\mu-}+\frac{1}{2} m_Z^2  Z^\mu Z_\mu\right)
\left(1+\frac{H^2+G^2}{v^2}+\frac{2H}{v}\right)
\nn
+{g^2\o2}G_+G_-\[W_+\cdot W_-+{\cos^22\theta_W\o2\cos^2\theta_W}Z^2\]
\nn+m_Z\((1+{H\o v})\p_\mu G-{G\o v}\p_\mu H\)Z^\mu
+i{g\cos2\theta_W\o2\cos\theta_W}Z_\mu\[G_-\na G_+-G_+\na G_-\]\nn
- {e\tan\theta_W\o2}Z_\mu\[(g(H+iG)+2m_W)W_+^\mu G_-+(g(H-iG)+2m_W)W_-^\mu G_+\]
\nn
+{ig\o2}[G_-W_{\mu+}\p^\mu (H+iG)-G_+W_{\mu-}\p^\mu (H-iG)]
\nn
+im_W \[( 1+{H-iG \o v})W_-\cdot\na G_+-( 1+{H+iG \o v})W_+\cdot\na G_-\]\ .
\label{higgs}
\eea

 It is worth noticing here that the interactions of the intermediate bosons $W_\pm$  with the electromagnetic field have two different natures. There are the interactions induced by the covariant derivatives, which are essentially the same for all charged fields, independently of their spin\footnote{In the $\xi=1$ gauge.}, which we call  {\it electric} interactions. There is however, in the second line of Eq. (\ref{iv}), a term proportional to the electromagnetic field tensor, which is typical of vector fields and should be called {\it magnetic} term. 

Then we must choose the gauge fixing part of the Lagrangian.

Our choice of the modified  't Hooft $\xi$-gauge fixing functions for the  intermediate vector bosons is:
\be {\cal G}_\pm=\na\cdot W_\pm\mp i \xi_W m_W G_\pm \quad\ ,\quad {\cal G}_Z=\p\cdot Z-\xi_Zm_Z G\quad\ ,\quad {\cal G}_A=\p\cdot A\ .\ee
What is new in the $\na$ operator replacing the ordinary derivative $\p\ .$
Introducing the multiplier fields $b_\pm\ ,\ b_Z$ and $b_A$ and the corresponding anti-ghosts $\bar c_\pm\ ,\ \bar c_Z$ and $\bar c_A$, the gauge fixing Lagrangian is:
\be\L_{\rm GF}={\cal S}\[ \bar c_-\({\cal G}_++ {\xi_W\o2} b_+ \)+\bar c_+\({\cal G}_-+ {\xi_W\o2} b_- \)+ \bar c_Z\({\cal G}_Z+{\xi_Zb_Z\o2}\)+\bar c_A\({\cal G}_A+{\a b_A \o2}\)\]\ ,\ee
where the action of the BRST operator ${\cal S}$ is given by
\bea&&{\cal S} W^\mu_\pm=\na^\mu c_\pm\mp ig\cos\theta_W(Z^\mu c_\pm- W_\pm^\mu c_Z)\pm ieW_\pm c_A\\&&
{\cal S}Z^\mu=\p^\mu c_Z-ig\cos\theta_W(W^\mu_+ c_--W^\mu_- c_+)\nn
{\cal S}A^\mu=\p^\mu c_A-ie(W^\mu_+ c_--W^\mu_- c_+)\nn
{\cal S}H={ig\o2}(c_-G_+-c_+G_-)+{m_Z\o v}c_ZG\nn
{\cal S}G={g\o2}(c_-G_++c_+G_-)-m_Zc_Z(1+{H\o v})\nn
{\cal S}G_\pm=im_W(1+{H\pm iG\o v})c_\pm\pm i({g\cos2\theta_W\o2\cos\theta_W}c_Z+ec_A)G_\pm \quad\ ,\quad{\cal S}\bar c_i=b_i\nn
{\cal S}c_\pm=\pm i (g\cos\theta_W+ec_A)c_\pm\quad\ ,\quad {\cal S}c_Z=ig\cos\theta_Wc_+c_-\quad\ ,\quad {\cal S}c_A=iec_+c_-\ .\nonumber\label{brst}\eea
After integration over the  multiplier fields $b_i$, one gets:
\be
\L_{\rm GF}=-{1\o\xi_W}{\cal G}_+{\cal G}_--{1\o2\xi_Z}{\cal G}_Z^2-{1\o2\a}{\cal G}_A^2-\bar c_-{\cal S}{\cal G}_+-\bar c_+{\cal S}{\cal G}_--\bar c_Z{\cal S}{\cal G}_Z-\bar c_A{\cal S}{\cal G}_A\ee
which can be written as the sum of two parts,
 a bosonic part:
\bea
\L_{\rm GF-B}&=&-{1\o\xi_W}\na\cdot W_+\na\cdot W_--\xi_W m_W^2G_+G_-+im_W\(G_+\na\cdot W_--G_-\na\cdot W_+\)\nn-{1\o2\xi_Z}(\p\cdot Z)^2-{\xi_Z m_Z^2\o2} G^2 +m_Z G \p\cdot Z-{1\o2\a}(\p\cdot A)^2\label{gfb}\eea and a Faddeev-Popov part:\eject
\bea
\L_{\rm GF-FP}&=&-\bar c_-\[\na^2 c_+ +\xi_W m_W^2\(1+{H+iG\o v}\)c_++e^2\(W_+\cdot W_-c_+-W_+^2 c_-\)\acca+iec_A \na\cdot W_++ig\cos\theta_W
\na\cdot\(W_+c_Z-Zc_+\)\acca+\xi_W m_WG_+({g\cos2\theta_W\o\cos\theta_W}c_Z+ec_A)\] +{\rm h.c.}\nn
-\bar c_Z\[\p^2 c_Z -ig\cos \theta_W\p\cdot(W_+c_--W_-c_+) +\xi_Zm^2_Z(1+{H\o v})c_Z\acca-{g\o2} \xi_Zm_Z(G_+c_-+G_-c_+)\] 
-\bar c_A\[\p^2 c_A -ie\p\cdot(W_+c_--W_-c_+)\] \ , \label{gffp} 
\eea where h.c. denotes the Hermitian conjugate of the former terms.

A particular consequence of our choice is  the loss of linearity of the gauge fixing functions ${\cal G}_\pm\ ,$ which follows the replacement  of ordinary derivatives with covariant ones. This  induces the systematic appearance of covariant derivatives without any further change in $\L_{\rm Higgs}$. On the contrary  in $\L_{\rm GF-FP}$ new couplings of the kind "$\bar c\ W^2 c$" appear,  as a typical consequence  of the mentioned non-linearity.

Notice that our gauge choice, besides avoiding vector-Goldstone boson mixing in the free Lagrangian, this being the goal of the 't Hooft choice, excludes vector-Goldstone boson transitions induced by the electromagnetic interactions. This is the main consequence of our choice.

For completeness we also give the propagators, which keep the usual form of the $\xi$-gauge propagators, that is,
for vector boson one has:
\be
\Delta_V^{\mu\nu}(k)={ g^{\mu\nu} +{\xi_V-1\o k^2-\xi_V m_V^2+i\epsilon}k^\mu k^\nu \o k^2-m_V^2+i\epsilon};
\qquad V=\gamma,W,Z,
\ee
with $m_\gamma=0$, $\xi_A=\a$, while scalar field propagators are given by:
\be
\Delta_s(k^2)=\frac{1}{m_s^2-k^2-i\epsilon}
\ee
with $m_{G^\pm}=m_{c^\pm}=\sqrt{\xi_W}m_W\ $, $m_G=m_{c^z}=\sqrt{\xi_Z}m_Z\ $, $m_{c^a}=0\ $.

\section{Renormalization}

We have now completed the  boring analysis of the new gauge fixed classical Lagrangian. It remains to discuss how the renormalization rules change. With this aim let us recall that, in the case of linear gauge fixing functions the renormalization of the theory is constrained, first of all, by the fulfillment of the Slavnov-Taylor identities. These are conveniently written, in the Rouet-Stora form \cite{rs} \cite{yz}, 
   as a linear functional differential equation  for the functional generator of the connected Green functions.  The same Slavnov-Taylor equations are translated into a bilinear, first order, functional differential equation for the proper, 1-particle irreducible, Green functional generator \cite{yz}.

A second family of constraints, typical of the linear gauge fixings, is given by the   equations of the multiplier fields  and of the anti-ghosts \cite{bbbc}.
 These equations, which correspond to linear, first order functional differential equations for both the connected and the proper Green functional generator, are not affected with renormalization corrections and constrain the parameters of the gauge fixing part of the Lagrangian. It remains to renormalize the gauge invariant part. This is done by field, mass and coupling constant renormalizations.
 
 If, on the contrary, the gauge fixing function is non-linear in the quantized fields, the gauge fixing parameters are not anymore protected from perturbative corrections, and hence further conditions are needed to complete the procedure. In other words the gauge fixing functions are composite operators which require renormalization prescriptions. In the present case the neutral vector field gauge fixing functions are linear, and hence the corresponding parameters are not renormalized.
 On the contrary the charged gauge field fixing functions are non-linear in the quantized field, due to the introduction of the electromagnetic covariant derivatives, therefore the corresponding parameters  must be renormalized. For example, there are divergent corrections to the  longitudinal part of the two point function of $W_\pm$ which must be subtracted by a renormalization prescription. Renormalizing the longitudinal part of this two point function is an essential step of the gauge fixing procedure.

 Furthermore, considering our gauge fixing choice, new terms are induced by the perturbative corrections into the gauge fixing functions ${\cal G}_\pm$.  In the present case perturbative corrections induce four new term into ${\cal G}_\pm$, these are:
 \be \a Z\cdot W_\pm+(\eta G\pm i\zeta H )G_\pm\pm i \bar c_\pm(\lambda c_A+\s c_Z)\ ,\label{gfflc}\ee where, assuming CP invariance, which is weakly broken by the fermionic corrections, $\eta\ ,\ \zeta\ ,\ \lambda$ and $\s$ are real parameters to be fixed by the renormalization rules. Notice that further ghost-anti-ghost terms, which would be allowed by power counting and charge conservation, are excluded by the field equations of the neutral anti-ghosts.
 
 The above comment means that, in the renormalization procedure, one has to take into account still more terms than those given  in Eq.s (\ref{iv}), (\ref{higgs}), (\ref{gfb}) and (\ref{gffp})  above.
 These are new terms in $\L_{\rm GF-B}$:
 \bea \xi_W\d\L_{\rm GF-B}&=&-(\eta^2G^2+\zeta^2 H^2-2\zeta\xi_Wm_WH)G_+G_-+\eta G(\na G_+\cdot W_-+\na G_-\cdot W_+)\nn
+\eta(\p G-\a GZ)\cdot (G_+W_-+G_-W_+)+i\zeta H(\na G_+\cdot W_--\na G_-\cdot W_+)\nn+i\[\zeta\p H-\a( \zeta H-\xi_Wm_W)Z\]\cdot (G_+W_--G_-W_+)
-a^2Z\cdot W_+Z\cdot W_-\nn
-\a(Z\cdot W_+\na\cdot W_-+Z\cdot W_-\na\cdot W_+)\ .\label{gfbv} \eea
Comparing this formula with Eq. (\ref{iv}) and Eq. (\ref{higgs}), notice that each single term in the sum is also present in the Higgs part of the invariant Lagrangian, that is, in $\L_{\rm Higgs}$. Therefore we see  that, with our gauge choice, the coefficients of these terms in $\L_{\rm Higgs}$ become explicitly gauge dependent. This does not break the Slavnov-Taylor identity.

New terms also appear in the Faddeev-Popov part of the Lagrangian:
\bea
\d\L_{\rm GF-FP}&=&\bar c_-\[ m_W\(\zeta H-i\eta G\)c_++(\eta m_Z-2\s m_W)G_+c_Z-2m_W\lambda G_+c_A-\a W_+\cdot\p c_Z\acca
+{2i\o\xi_W}\na\cdot W_+(\lambda c_A+\s c_Z)+ig\a\cos\theta_W (W_+^2c_--W_+\cdot W_-c_+)-\a Z\cdot\na c_+\acca
+{g\o2}\zeta H^2 c_++i\a g \cos\theta_WZ^2 c_+-i\a W_+\cdot Z \((g\cos\theta_W+{\s\o\xi_W})c_Z+(e+{\lambda\o\xi })c_A \)\acca
+{g\o2}(\zeta-\eta)G_+^2 c_-
+\(e-{2\lambda\o\xi_W}\)\(\zeta H-i\eta G\)G_+c_A\acca
 +{g\o2}\(\eta G^2 +i(\zeta-\eta)HG-(\zeta+\eta)G_+G_-\)c_+ \acca
 -\({g\o2\cos\theta_W}(\eta+\cos2\theta_W\zeta)+{2\zeta\s \o\xi_W}\)HG_+c_Z\acca
 +i\({g\o2\cos\theta_W}(\zeta+\cos2\theta_W\eta)+{2\eta\s\o\xi_W}\)GG_+c_Z\] +{\rm h.c.}  \nn
+2(\lambda e+\s g\cos\theta_W)\bar c_+\bar c_-c_+ c_-\ . \label{gffpv}
\eea  Here the  terms in the first two lines modify the coefficients of the same terms in $\L_{\rm GF-FP}$, while the rest accounts for further four field couplings. Notice that, among these new  four field couplings, only those in the third line and their Hermitian conjugate, involve two physical fields, while the other couplings involve, at least three unphysical ones.

Thus, considering the  renormalization effects of the non linear terms in the gauge fixing functions ${\cal G}_\pm$, we see that the higher loop corrections, besides acting in the coefficients of ${\cal L}_V\ ,\ {\cal L}_{Higgs}\ ,\ {\cal L}_{GF}$ and ${\cal L}_{GF-FP}$, and hence inducing higher loop corrections to the coefficients of the relevant diagrams, generate new diagrams through the mentioned new, four field, vertices. With the exceptions of those contained in the third line of Eq. (\ref{gffpv}),  the contributions of the new diagrams to the physical amplitudes do not appear before the third loop order. Indeed the diagrams must contain at least two unphysical particle loops and one coupling constant induced by the  one loop corrections. On the contrary the first loop correction effects are foreseen at two loops in processes such as the Higgs boson elastic scattering with photons, or else with $W$'s and $Z$'s.

This exhausts the list of drawbacks in the renormalization of our scheme, which are due to the electromagnetic gauge invariant choice. It must be clear, however, that, if one choses at the classical level $\xi_V\equiv 1$ and $\eta=\zeta=0\ ,$ the Feynman rules appear greatly simplified and, as above discussed,  only the calculation of higher loop corrections to the transition amplitudes oblige us  to deal with the terms appearing in   Eq.s (\ref{gfbv}) and (\ref{gffpv}).

Considering instead the advantages of our scheme, these are consequences of the QED Ward identities and of the $\xi$-gauge choice. The content of the QED Ward identities is the invariance under electromagnetic gauge transformations of  the proper Green functional generator, which  accounts for the contribution of all the loop diagrams and hence plays the role of the effective action of the theory, deprived of  the Feynman gauge fixing term.
If we denote this proper generator by the functional $\Gamma[A_\mu,\Phi_\pm, \Phi_0]$, where we have collectively denoted by $\Phi_\pm$ the charged fields and sources, and by $\Phi_0$ the neutral ones (excluding the electromagnetic field $A$) we have the functional differential equation\footnote{Which is consistent with the Slavnov Taylor identity with the further constraint of the electromagnetic ghost field equation.}:
\be \p_\mu {\d\Gamma\o\d A_\mu(x)}-ie\sum\Phi_+(x){\d\Gamma\o\d \Phi_+(x)}+ie\sum\Phi_-(x){\d\Gamma\o\d \Phi_-(x)}={1\o\a}\p^2(\p\cdot A(x))\ ,\label{ewi}\ee whose right-hand side term gives the variation of the Feynman gauge fixing term.
The first consequence of  this equation  is the non-renormalization of the field charges which explicitly appear in the identity, in much the same way as the parameter $\a$. The same non-renormalization properties can be  proved  also in the framework of linear gauge choices, but using a much less direct argument \cite{bbbc}. 

Let us consider now how the Feynman diagrams get modified in our scheme. As already mentioned, commenting Eq.'s (\ref{iv}) and (\ref{higgs}), among  the electromagnetic interactions of the charged fields,  we have identified the, so called, electric interactions and magnetic ones, which do not interrupt the charged field lines, the same is true for the interactions of the form
$AZ\Phi_+\Phi_-$ which appear in the third line of Eq. (\ref{iv}) and in the fifth line of Eq. (\ref{higgs}). 
The vertices corresponding to these interactions appear in a diagram as vertices where two propagators of the same charged field join together. This suggest the introduction of the idea of generalized charged particle propagators, which are identified with  chains of propagators of the same charged field joined by the mentioned vertices.
There are however interaction terms in the last three lines of Eq. (\ref{higgs}) and in the second, fourth and fifth lines of Eq. (\ref{gffp}), joining different generalized-propagators, e. g. that of $W_\pm$ with that of $G_\pm$, the corresponding vertices are end points of generalized propagators.

The Feynman diagrams contributing to a given process can be grouped together when they only differ in the insertion of their photon lines, assumed in the same number. Thus a group of diagrams is identified by a common sub diagram made up of generalized and neutral particle propagators with photon lines connecting these propagators in all possible ways.

The sum of all the diagrams belonging to the same group gives a homogeneous contribution to Eq. (\ref{ewi}). This, in particular, reduces the divergence degree of the sum of the diagrams of the same group. The above described diagram structure shows that our choice strongly reduces the number of groups of diagrams, indeed in the linear $\xi$-gauges there are three line electromagnetic vertices joining different charged particle propagators which do not appear in our scheme. In the example of the Higgs into two photons decay this reduction is apparent. 

Last, but not least, we have already mentioned the crucial role of the QED Ward identities in the analysis of the low energy electromagnetic radiative correction to the physical processes. Concerning this point one should, however, take into account that, as is well known \cite{yfs}, at least at the leading order in $\log (\Delta E/E)$, where $\Delta E$ is the energy resolution of the electromagnetic detector and $E$  is the energy of the process,  the radiative corrections correspond to characteristic factors multiplying
the hard process cross sections, which depend on the hard-soft photon separation energy. Now the  hard process cross sections are gauge fixing independent, and hence it is natural to compute them with the simplest possible gauge fixing choice. What remains to compute consists in the radiative correction factors which depend on the nature of the initial and final charged particles. These are typically charged spinors whose electromagnetic interactions do not depend on the electroweak gauge choice. Thus the role of our scheme in the evaluations  of the leading order radiative corrections  amounts to guaranteeing that the usual computing methods are well founded. Considering sub leading terms our scheme might play a more relevant role.

\section{Conclusions}

With few fundamental exceptions, e. g. the $\xi$ gauges of Feynman and 't Hooft, the background gauges and some non-covariant choices, often the analyses of new gauge fixing choices  are a often of a merely mainly academic nature. This paper, written  on the occasion of Andrey's seventy fifth birthday, with the aim of clarifying the relations between one Andrey's major achievements, namely the Slavnov-Taylor identities, with the old Ward identities of QED, tries to convince the reader that an electromagnetic covariant gauge choice, inspired by the background choices, might be useful.

 We have tried to honestly list the drawbacks of the suggested method;  these are essentially due to the renormalization effects of the non-linear gauge choice, which rapidly grow  with the weak perturbative order of the calculations. The  QED Ward identities are the main source of the advantages of the method over more standard   choices. These advantages mainly appear in the calculation of electromagnetic processes. Typically one finds a reduction of the ultra-violet degrees of diagrams and also of their number.
Of course,  a particular role is played by our method in the calculation of electromagnetic radiative corrections. 

Let us finally notice that some particular gauge fixing choices belonging to the class described above have been made e.g. in the references listed in \cite{rxi}. I thank S. Inoue for his help on this point.
 



\end{document}